\documentclass[a4paper,12pt]{article}
\usepackage[dvips]{graphicx}\usepackage{epsfig}\usepackage{amssymb}
\usepackage{amsfonts}\usepackage{hhline}
 \usepackage{amsmath}   \usepackage{pifont}
\def\ea{{\it et al.\,}} \def\eg{{\it e.g.,\,}} \def\ie{{\it i.e.,\,}}
\def\eV{{\rm e\kern-0.12em V}} \def\GeV{{\rm G}\eV} \def\MeV{{\rm M}\eV}
\def\albar{\relax\ifmmode{\bar{\alpha}}\else{$\,\bar{\alpha}${ }}\fi}
\def\gbar{\relax\ifmmode{\bar{g}}\else{$\,\bar{g}${ }}\fi}
\def\fbar{\relax\ifmmode{\bar{f}}\else{$\,\bar{f}${ }}\fi}
\def\albarQ{\relax\ifmmode{\bar{\alpha}(Q^2)}\else{$\albar(Q^2)${ }}\fi}
\def\albarsf{\relax\ifmmode{\bar{\alpha}_{\rm SF}}\else{$\,\bar{\alpha}_{\rm SF}${ }}\fi}
\def\alphasf{\relax\ifmmode{\alpha_{\rm SF}}\else{$\,\alpha_{\rm SF}${ }}\fi}
\def\albars{\relax\ifmmode{\bar{\alpha}_s}\else{$\,\bar{\alpha}_s${ }}\fi}
\def\as{\relax\ifmmode\,\alpha_s\,\else{$\,\alpha_s\,${ }}\fi}
\def\asQ{\relax\ifmmode{\bar{\alpha}_s(Q^2)}\else{$\albars(Q^2)${ }}\fi}
\newcommand{\beglab}{\begin{equation}\label}
\newcommand{\beq}{\begin{equation}} \newcommand{\eeq}{\end{equation}}
\newcommand{\ba}{\begin{eqnarray}}
\begin{document}
\begin{flushright} {
JINR \ Preprint \ E2-2003--1 }
\end{flushright}
\vspace{8mm}

\begin{center}
{\large\bf On \ the \ Fourier \ transformation \ of \\
        Renormalization Invariant Coupling }\\

{\large D.V.Shirkov }\\

\vspace{1mm}
{\it Bogoliubov Lab. of Theor. Phys., JINR, Dubna, 141980, Russia}\\
{\it shirkovd@thsun1.jinr.ru} \\
\end{center}
\begin{abstract}
Integral transformations of the QCD invariant (running)
coupling and of some related objects are discussed. Special attention
is paid to the Fourier transformation, that is to transition from the
space-time to the energy--momentum representation.\par
 The conclusion is that the condition of possibility of such a transition
provides us with one more argument against the real existence of
unphysical singularities observed in the perturbative QCD. \par
  The second conclusion relates to the way of ``translation" of some
singular long--range asymptotic behaviors to the infrared momentum region.
Such a transition has to be performed with the due account of the
Tauberian theorem. This comment relates to the recent ALPHA collaboration
results on the asymptotic behavior of the QCD effective coupling
obtained by  lattice simulation.
 \end{abstract}
\tableofcontents

\section{Invariant coupling in different representations}

 Current practice of the QFT calculations, as a rule, employs expression
of observables and of some other intermediate renormalization--group (RG)
invariant, or covariant, objects in terms of the {\sf invariant coupling}
function (also referred to as the ``invariant, or running coupling
constant"\footnote{In view of semantic absurdity of the last term, we use
the expression ``invariant coupling function" or {\sl ``invariant coupling"
\/}.}). This invariant coupling $\,\albar(Q^2)\,$ is a real function of a
real positive argument $\,Q^2\equiv {\bf Q}^2-Q_0^2\,,$ momentum transfer
squared.\par

The notion of invariant coupling (IC) has initially been
introduced\footnote{Under the name of ``invariant charge" (of electron), the
natural one in the QED context.} --- see, e.g., refs.\cite{rg56,kniga} on
renormalization group --- in terms of a product of {\it real} constants
$\,z_i\,$ entering into finite Dyson renormalization transformations
\beglab{dyson-tr}
D_i(... , \alpha) \to D'_i(... , \alpha')=z_i^{-1}\, D_i(... ,
z_3\,\alpha)\,;\quad\alpha \to \alpha'= z_3\,\alpha\,,\eeq
with particle propagators $\/D_i(... ,\alpha)= D_i(Q, m,\mu;\alpha)\,$
taken in the energy--momentum representation. Here, the IC is expressed
in terms of scalar (\ie, Lorentz--invariant) QFT amplitudes taken also in
the momentum representation. For instance, in QED one has
$$
\albarQ\equiv\alpha d\left(\frac{Q^2}{\mu^2},\frac{m^2}{\mu^2};
\alpha\right)\,$$
with $\/d\,,$ the transverse photon propagator amplitude which is real
in the Euclidean domain. In QCD, \ IC \ \asQ \ is usually defined in a
similar way as a product of the expansion parameter \as with an
appropriate scalar vertex and propagator amplitudes, see, e.g.,
\cite{gw73} and references therein. \par

  This IC is commonly used for parameterization of the
renormalization--invariant quantities, in particular, of observables. To
this goal, one should take an invariant quantity of interest in the
appropriate representation. In particular, in terms of \asQ there could be
presented only an object taken in the momentum representation. Moreover,
this object to be expressible in terms of the real function \asQ should be
real itself. In the QFT case, this corresponds to the space--like Euclidean
domain with $Q^2>0\,.$\par

 Meanwhile, some observables, like effective cross--sections, are functions
of another Lorentz--in\-va\-riant argument $s\,,$ the center of mass energy
squared. As it is well known, the polarization operator $\Pi(Q^2)\,,$
being represented in the form of the Cauchy type integral, provides us with
a technical means for the integral connection of invariant functions of the
$Q^2\,$ and $s\,$ arguments. Its logarithmic derivative, known as the
Adler function, is defined via integral
\beglab{adler}                                                       
D(Q^2) = Q^2\int^{\infty}_0 \frac{d s}{(s+Q^2)^2}\,R(s)\,\,,\eeq
close to the K\"all\'en--Lehmann spectral representation for $\Pi(Q^2)\,.$
Here, $R(s)\,$ being the imaginary part of the polarization operator, due
to the optical theorem, is proportional to the total cross--section which
is an observable quantity. \par

Further on, we shall take relation (\ref{adler}) as the base for definition
of the integral operation $\mathbb{D}\,$ transforming a real function
$\,M(s)\,$ of the positive (time--like) argument into another real function
$\,E(Q^2)\,$ of the positive real (space--like) argument $Q^2\,$ ---
see below eq.(\ref{Dtrans}). \par
   More generally, we consider a linear integral transformation with kernel
 $K(x)\,,$ depending on one argument
\beglab{lin-gen}
f(x) \to F(y)=\mathbb{L}\,[f](y)=\int_0^\infty d(xy)\, K(xy)\, f(x)\eeq                
that comprises various kinds of one-dimensional Fourier transformations
(usual, Laplace, cosine--Fourier, sine--Fourier and some others, \eg with a
kernel in the form of the Bessel function that could be derived from the
four--dimensional Fourier one) and, with $K(z)=\left(1+z^2\right)^2,$ just
mentioned the ``Adler transformation" $\mathbb{D}\,$ defined below by
eq.(\ref{Dtrans}).\medskip

  In Section 2, we start with the discussion of the interrelation between
the RG invariance and integral transformation. Our first explicit example
taken from the so--called Analytic Perturbation Theory (APT) concerns a
transition from the Minkowskian energy--momentum domain to the
corresponding Euclidean one. We continue with the transition from the
``distance" (\ie, three--dimensional space) representation to the common
momentum (transfer) one using the one--dimensional sine--Fourier
transformation.\par

  Further on, in Section 3, we shall pay special attention to the relation
between a {\it long--range distance} and {\it infrared momentum}
asymptotic behaviors in QCD, and comment on the interpretation of
some recent results of the ALPHA collaboration.

\section{Integral transformations \label{s2}}
\subsection{RG invariance through integral transformation\label{ss2.1}}

 First of all, note that transformation $\mathbb{L}\,,$ eq.(3), is compatible
 with the RG transformation.
For example, in the massless case, let some initial function
$f(x/\mu;\alpha)\,$ be invariant with respect to the RG transformation
\beglab{rg-tr}
R_\tau: \left\{\mu^2\to (\mu')^2=\mu^2\,\tau \,;\quad\alpha\to
\alpha'=\albar_{\rm tr}(\tau, \alpha)\right\}\,. \eeq
Then, its integral image $\,F(y \mu;\alpha)\,=\mathbb{L}[f](y)\,$ will be
invariant with respect to the same RG transformation (\ref{rg-tr}) --- and
vice versa --- with {\it the same transformation function\/}
$\albar_{\rm tr}\,$ which satisfies the functional equation
\beglab{rg-fe}
\albar(\theta\tau,\alpha)=\albar(\theta,\albar(\tau,\alpha))\,\eeq
that follows from the group composition law $R_{\theta}R_\tau=
R_{\theta\tau}\,.$  Note here, that eq.(\ref{rg-fe}) implies the
``canonical normalization condition" for the transformation function
 $\albar_{\rm tr}$
 \beglab{canon}
\albar(1,\alpha)=\alpha\, \eeq
that corresponds to the identity transformation $R_1\,.$

 Relation (\ref{adler}) is commonly used to define the Adler function
$D\,$. However, we shall treat it instead as a definition of the integral
operation $\mathbb{D}\,$ transforming a function $\,M(s)\,$ of the
positive real argument $\,s\,$ into another function $\,E(Q^2)\,$ of the
positive real argument $Q^2\,$
\beglab{Dtrans}
M(s) \to\mathbb{D}[M](Q^2)\equiv Q^2\int^{\infty}_0 \frac{d s}
{(s+Q^2)^2}\,M(s)\, = E(Q^2)\, \eeq                                 
with the reverse operation $\mathbb{R}\,=[\mathbb{D}]^{-1}$ that can
be expressed in terms of the contour integral; for more details about
these operations, see, \eg  ref.\cite{pr00-46,tmp01}.\par

 The transformations $\mathbb{D}\,$ and $\mathbb{R}\,$ preserve the RG
invariance; for instance, they transform a renormalization invariant function
$M(s/\mu^2;\alpha)\,$ into another RG invariant $E(Q^2/\mu^2;\alpha)\,$
and vice versa. The first of these invariants $M\,$ should be representable
as a function of an adequate ``Minkowskian invariant coupling" $\alpha(s)\,,$
while the other $E\,$ --- via the common invariant coupling $\albarQ\,.$
In particular, this means that starting with the usual QCD coupling \asQ,
by the operation $\mathbb{R}\,$ it is possible to define \footnote{
However, in practice, the nonphysical singularity in perturbative \asQ
prevents one from a straightforward performance of the integration
procedure. In more detail, this issue has been discussed in \cite{tmp01}.}
the QCD effective coupling in the time--like Minkowskian region
$\alpha_M(s)=\mathbb{R}[\albars](s)\,$ in terms of \asQ, a QCD invariant
coupling initially introduced in the Euclidean domain. \par \smallskip

 Another example is related to the physical amplitude $A\,$ depending on a
couple of Lorentz invariant kinematic arguments $s\,$ and $t\,.$ Let it be
renormalization invariant
$$
A(s,t;\mu,\alpha)=A(s,t;\mu',\alpha')\,.$$
  Here, there are two possibilities. The first one deals with an integral
transformation with respect to one of the two kinematic arguments like an
eikonal transformation $F(s,t)\to\Phi(s,b)\,.$ A new amplitude $\Phi\,$
has the RG transformation property similar to that of $F\,.$ Like in the
general case of several kinematic arguments, the RG transformation
properties of $F$ and $\Phi\,$ are not very useful. \par

 However, this situation changes for an integral transformation of the
second type involving some function of ratio of both the variables
$\varphi(s/t)\,.$ Such a transformation
$$
A(s,t) \to A_k(s)=\int_0^\infty A(s,t)\,K_{(k)}(s/t)\,
d\varphi(s/t)\,;\quad k=0,1,2,\dots\,.$$
projects the initial function of two (kinematic) arguments onto a (set of)
function(s) $A_k(s)\,$ of one argument which is(are) RG invariant. The
transition to partial waves and to moments of structure functions provides
us with examples. \medskip

 One more example of an integral transformation of the ``one--argument"
function compatible with the RG invariance is given by the transformation
of the Fourier type which follows the general linear form,
eq.(\ref{lin-gen}). It relates the function $f(r)\,$ of the space-time
Lorentz invariant argument $\,r=\sqrt{{\bf r}^2-t^2}\,$ with the function
$\,F(Q)\,$ of an energy--momentum invariant argument
$\,Q=\sqrt{Q^2={\bf Q}^2-Q_0^2}\,.$ \medskip

 As it is well known, the Dyson transformation (\ref{dyson-tr}), being ``a
self-similarity property of the Schwinger--Dyson equations"\cite{bbbsh57},
can be, on equal footing, considered in both the (energy-)momentum --- like
in eq.(\ref{dyson-tr}) --- and (time-) space representation, that is for
functions like
$$
D(...,\alpha)=\tilde{D}(x,m,\mu;\alpha)\,;\: x=\{ {\bf r}, t\} .$$

In the QED context, this latter picture\footnote{We shall refer to it
as to the ``distance representation".} was used by Dirac while discussing
polarization of vacuum in terms of {\it effective charge of electron}
first introduced by him\cite{dirac34} as a space distribution
\beglab{dirac}
\bar{e}_D(r)=e_0\left\{1-\frac{\alpha_0}{3\pi}
\ln\frac{r}{r_e}\right\}\,;\quad\alpha_0=\frac{e_0^2}{4\pi}
\simeq \frac{1}{137.0360}\quad r_e=\frac{h}{m_ec} \eeq                
of the electric charge around the point ``bare" electron.\par

 More recently, an analogous object, the QCD effective coupling
$\gbar^2(L)\,,$ which is a function of the spatial size $L\,$ of a lattice,
has been introduced\cite{lu92} and used\cite{alf95} -- \cite{alf01b} by
ALPHA collaboration. \par

 Some other examples of the relation between integral transformations
and renormalization group symmetries in problems of classical and
mathematical physics can be found in a fresh review paper \cite{kov02}
(see, especially, Example 2 on page 358 and references therein). \par

 The issue of correlation between the RG formulations in different pictures
has two aspects. The first one deals with the basic notions and objects of
RG transformations. The second one is that of these objects correlation via
an appropriate integral transformation. \par

  According to the terminology formulated in Refs.\cite{kniga2,dv02}, an
invariant function satisfying the functional equation (\ref{rg-fe}) and
canonical normalization condition should be named {\it effective coupling\/}
(EC). To the case of an invariant function $\albar_N\,$ with a more general
normalization
 $$
 \albar_N(1,\alpha)=N(\alpha)\neq\alpha\,$$
there corresponds a term {\it invariant coupling\/} (IC). Contrary to EC,
it can not be used as a function transforming a coupling constant in
(\ref{rg-tr}). Generally, integral transformation (\ref{lin-gen}) maps an
EC $\albar(x)\,$ onto some other RG--invariant function $A(y)\,$ which is
IC rather than EC. \par\medskip

  A few questions are in order:\par
a) How to define $\albar\,$ and its integral image $A\,$ explicitly
for  the given QFT model?\par
b) How to relate them? \par
c) Which of them can be used as the transformation function
$\alpha_{\rm tr}\,?$\par   \medskip

  ``Immediate" answers --- \par
a] Use a common algorithm with perturbative beta--functions to
define $\albar\,$ and $A\,$, \par
b] Relate them by an appropriate transformation (\ref{lin-gen}) \\
---  turn out to be incompatible with the each other. \par \smallskip

\subsection{From the c.m. energy to the momentum picture \label{ss2.2}}
  This situation can be illuminated by the known answers to the same
questions for the integral transformation (\ref{Dtrans}). Let us write it
down for the particular case of the coupling function transformation
\beglab{d-tral}
\alpha(s) \to\mathbb{D}[\alpha](Q^2)\equiv Q^2\int^{\infty}_0
\frac{d s}{(s+Q^2)^2}\,\cdot \alpha(s)\,= \albarQ\, .\eeq             
It turns out to be impossible to use here a common perturbative QCD coupling
\asQ as $\albar(Q^2)\,$ because of its unphysical singularities, like the
Landau pole at $Q^2=\Lambda^2\,,$ which contradicts the integral expression
(\ref{d-tral}). The latter implies that function \albarQ should be free of
any singularities outside a cut $0>Q^2>-\infty\,.$ The same is true for the
candidature of \asQ for the role of function $\alpha(s)\,.$ One of the
possible solutions that was proposed in the so-called Analytic Perturbation
Theory (APT) (see, refs. \cite{ms97,tmp01}) consists in using of \asQ only
as a prototype for both the $\albar(Q^2)\,$ and $\alpha(s)\,$ which in the
one--loop case with
\beglab{als1}
\albars^{(1)}(Q^2)=\frac{1}{\beta_0\ln(Q^2/\Lambda^2)}\,;\quad
\Lambda^2=\mu^2 \exp\left(-\frac{1}{\beta_0 \as}\right)\eeq      
turn out to be
\beglab{ic5}
\alpha(s)= \left.\frac{1}{\pi\beta_0}\arccos\frac{L_s}{\sqrt{L_s^2
+\pi^2}}\right|_{L_s>0}=\frac{\arctan(\pi/L_s)}{\pi\beta_0}\,;
\quad L_s=\ln\frac{s}{\Lambda^2}\,,\eeq                             
and
\beglab{alphan}
\albar(Q^2)=\frac{1}{\beta_0}\left[\frac{1}{L}-\frac{\Lambda^2}{Q^2-
\Lambda^2}\right]\,;\quad L=\ln\frac{Q^2}{\Lambda^2}\,.\eeq        

  Both these two functions, corresponding to (\ref{als1}) in the weak
coupling limit, are ghost--free monotonous functions related by the
transformation (\ref{d-tral}) $\albar(Q^2)=\mathbb{D}[\alpha](Q^2)\,$
and its reverse. \par

  As it has been noticed in Ref.\cite{tmp01}, transitions from expression
(\ref{als1}) for $\albars^{(1)}(Q^2)\,$ to $\alpha(s)\,$ and $\albarQ\,$
can be represented as a consequence of a transition from the usual QCD
coupling constant $\as\,$ to the new ones
$$
\as\to\alpha_M=\frac{1}{\pi\beta_0}\arccos\frac{1}{\sqrt{1+\pi^2
\beta_0^2\as^2}}\,=\frac{1}{\pi\beta_0}\arctan(\pi\beta_0\as)\,,$$
and
$$
\as\to\alpha_E=\as+\frac{1}{\beta_0}\left(1-e^{1/\beta_0
\as}\right)^{-1}\,.$$
Hence, a transition from the Minkowskian coupling $\alpha(s)\,$ to the
Euclidean one $\albarQ\,$ is equivalent to that one induced by the
following coupling constant transformation :
\beglab{am-ae}
\alpha_M\to\alpha_E(\alpha_M)=\frac{1}{\pi\beta_0}\tan(\pi\beta_0\alpha_M)
+\frac{1}{\beta_0}\cdot\frac{1}{1-e^{\pi\cot(\pi\beta_0\alpha_M)}}\,\eeq
at $0<\alpha_M\leq 1/\beta_0 \,$ and $0 <\alpha_E\leq 1/\beta_0\,.$ \par
In turn, this implies that the integral transformation (\ref{d-tral}),
generally changes the normalization of coupling functions and, in
particular, maps the EC onto IC.

\subsection{Momentum and distance representations \label{ss2.3}}
 Turn now to the Fourier transformation. It will be convenient to use the
modified sine--Fourier transformation\footnote{that is a particular case of
(\ref{lin-gen}) with $K(z)=2\sin(z)/\pi z\,$ and follows from the usual
3-dimensional one
$$
\,\bar{\psi}(Q)=(2\pi)^{-2}\int d{\bf r} \psi(r)e^{i{\bf Q r}}\quad
\mbox{with}\quad F(Q)=Q^2\bar{\psi}(Q)\,,\quad f(r)=r\,\psi(r)\,.$$}
of the RG--compatible form
\beglab{f-sin}
\mathbb{F}_{\rm sin/r}[f](Q)\equiv\frac{2}{\pi}
\,\int^\infty_0\frac{d\,r}{r}\,\sin(Qr)\,f(r)=F(Q)\,\eeq      
and its reverse
$$
\mathbb{F}^{-1}_{\rm sin/r}[F](r)\equiv r\,\int^\infty_0\,d Q\,\sin(Qr)
\,F(Q)= f(r)\,.$$

 Now, rewrite expression (\ref{dirac}) squared in a more contemporary and
general notation
\beglab{dir-mod}
\alpha_D^{pt}\left(\frac{r^2}{r^2_{\mu}},\alpha_{\mu}\right)=\alpha_{\mu}
\left\{1-\frac{\alpha_{\mu}}{3\pi}\ln\frac{r^2}{r^2_{\mu}}\right\}\,\eeq
($r_{\mu}\,$ being a reference distance) with the correspondence
relation $\alpha_{\mu}=\alpha_0\,$ at $r_{\mu}=r_e\,.$  This {\it
distance--representation perturbative EC\/}
$\,\alpha_D^{pt}(\rho^2,\alpha)\,$ can be connected by the transformation
$$
\albar\left(Q^2/\mu^2,\alpha\right)=\mathbb{F}_{\rm sin/r}[\alpha_D^{pt}]
(Q^2)\,$$
with a more common perturbative QED coupling in the momentum
representation
\beglab{albar-q}
\albar_{pt}\left(\frac{Q^2}{\mu^2},\alpha_{\mu}\right)=\alpha_{\mu}
\left\{1+\frac{\alpha_{\mu}}{3\pi}\ln\frac{Q^2}{\mu^2}\right\}\,
\quad \mbox{with}\quad\mu=\frac{c_E}{r_{\mu}}\,\eeq             
and $c_E=e^{-\,\mathbf{C}}=0.5614\,.$ Expression (\ref{albar-q}) can
also be obtained from (\ref{dir-mod}) by the argument substitution
\beglab{qscale}
r\to c_E/Q\,\,,\eeq                                             
 that is close to the quantum--mechanical correspondence relation
$r\to 1/Q\,$ being slightly modified by changing the $Q\,$ scale. By the
way, the positive effect of the $c_F\,$ factor is clearly seen in our
Figure 2 --- see below.\smallskip

 Quite analogously with the presented example, both the functions
$\albar^{RG}\,$ and $\alpha_D^{RG}\,,$ being explicitly defined (in the
familiar form of the geometric progression sum) by the common RG
perturbation--based procedure with (\ref{dir-mod}) and (\ref{albar-q}) as
inputs, are not related by eq.(\ref{f-sin}). Instead, they are connected by
a more involved relation $\albar^{RG}(Q)=\Psi\left\{\mathbb{F}_{\rm sin}
\left[\alpha_D^{RG}\,\right](Q)\right\}\,.$\par

 Moreover, generally, Fourier transformation maps an effective coupling
onto some invariant coupling that cannot be used as $\alpha_{\rm tr}\,$ in
(\ref{rg-fe}). For example, if one starts with the perturbative
distance--representation EC of the form
\beglab{dir-mod2}
\alpha_D^{pt}\left(\rho^2,\alpha\right)=\alpha-\beta_0(\alpha)
\ \ln\rho^2\,+\alpha\left(\frac{\alpha}{3\pi}\right)^2\ln^2\rho^2\,
;\quad\rho^2 =\frac{r^2}{r^2_{\mu}};\quad \alpha=\alpha_\mu\eeq         
then, as a result of the sine-Fourier transformation (\ref{f-sin}), one
arrives at
\beglab{lambda-e}
\albar_{pt}\left(q^2,\alpha\right)=\alpha+\beta(\alpha)\ln q^2+\alpha
\left(\frac{\alpha}{3\pi}\right)^2\left\{\ln^2 q^2+\Delta_2\right\}\,\eeq 
with
$$
\quad q^2=\frac{Q^2}{\mu^2}=\left(\frac{Qr_\mu}{c_E}\right)^2\,;
\quad \Delta_2=\frac{\pi^2}{12}\,.$$
 Hence, $\albar(1,\alpha)=\alpha+\alpha\left(\alpha/3\pi\right)^2\Delta_2
=\alpha'\neq\alpha\,$ and $\albar(q^2, \alpha)\,$ is an invariant
coupling\footnote{Meanwhile, this feature is not essential (like for
transformation (\ref{am-ae})) in the weak coupling case at the one-- and
two--loop levels.}. By transition to another coupling constant
$\alpha\to\alpha'\,$ it can be transformed (see Section 3 in \cite{dv02})
into the EC $\albar'(q^2,\alpha')\,.$ The transformation is relevant to
the $\Psi(\alpha)\,$ dependence. This could be important at the strong
coupling case in the IR region. \par

  Note also that the logarithmic terms in both (\ref{dir-mod2}) and
(\ref{lambda-e}) yield the sums of geometric progression with arguments of
logarithms related by eq.(\ref{qscale}).

\section{Long--range and infrared asymptotics\label{s3}}
\subsection{Tauberian theorem   \label{ss3.1}}

Turn now to the particular issue of correlation between asymptotic
behaviors of the functions $f\,$ and $\bar{f}\,$ related by the Fourier
transformation.\par
  This correlation is popular in quantum mechanics where, quite often,
one uses the so--called ``quantum--mechanical correspondence relation"
$$
\,r\to 1/Q\,,\eqno{(QMC)}$$
which in the IR case is equivalent to
\beglab{qm-corr}
F(Q)\sim f(Q^{-1})\quad \mbox{as}\quad Q\to 0\,.\eeq            

  Heuristically, this last feature could be simply understood by a change of
the integration variable $r \to x=rQ\,$ in the general linear transformation
(\ref{lin-gen})
$$
F(Q)=\int^\infty_0\frac{d\,x}{x}\,K(x)\,\,
f\left(\frac{x}{Q}\right)\,.\eqno{(3a)} $$

 However, for a more rigorous derivation of (\ref{qm-corr}) one needs to
specify some asymptotic property of the function $f(r)\,$ as
$r\to\infty\,.$  In short, this can be formulated as the {\sf Tauberian
theorem}\footnote{Originally, under the
name of Tauberian theorems one implied statements concerning the relation
between {\it summability} and {\it convergence} of series. More recently,
in the middle of XX century, this term started to be used in the context
of asymptotic properties of integral transformations. Here, we give only
a crude outline of this important theorem for the Fourier transformation,
the sketch that is sufficient for our application. For a more complete
and rigorous exposition of this matter the reader is referred to
refs.\cite{post80,vdz86}.}:
( Here, the symbol $``\sim"\,$ means ``behaves like".)
\begin{quote}
{\footnotesize\it If function $\,f(r)\,$ asymptotically satisfies
``the separability condition"
$$
f(kr)\sim C\phi(r)\,\varrho(k)\quad\mbox{as}\quad k\to \infty\,
\quad\mbox{with}\quad C\neq 0\,,                              \eqno(S)$$
then, under some additional conditions, its Fourier image obeys the property
$$F(Q)\sim\varrho(1/Q)\,\quad\mbox{as}\quad Q\to 0\,,         \eqno(T) $$}
\end{quote}
--- that, with some reservation, follows from eq.(3a).

Now, for a definite class of functions $\varrho(k)\,$ entering into
the condition (S), \eg of power and/or logarithmic type
\beglab{tauber}
f(r)\sim \varrho(r) \sim r^\beta\,(\ln r)^\gamma\,(\ln\ln r)^\delta
\dots\,\quad\mbox{as} \quad r\to\infty\,, \eeq
it is possible to obtain from (T) the correspondence rule (\ref{qm-corr}).\par

\subsection{Illustrations \label{ss3.2}}

 Consider few examples given in the Table.
\begin{center}
{\sf T a b l e} \renewcommand{\arraystretch}{1.25}\smallskip

\begin{tabular}{|c||c|c||c|c|}  \hline
 & $f(r)$ & $F(Q)=\mathbb{F}_{\rm sin/r}[f]$&$S^f_{\rm LR}$&
 {\small $(20)_{\rm IR}$} \\ \hhline{|=#=|=#-|-|} 
\{*\}& $r F(r)$ &{\small $2f(Q)/\pi Q$} &$r\to\infty$
&$Q\to 0$  \\ \hhline{|-#-|-#=|=|}
\{1\}&$r^\nu$;\,{\scriptsize $0\leq\nu<1$}&{\small $\frac{2}{\pi\nu}\sin
\left(\frac{\pi\nu}{2}\right)\Gamma(1+\nu)\:Q^{-\nu}$}&+&+ \\ \hline
\{2\}&~$\ln r$ &{\small $\ln Q^{-1}-\mathbf{C}\,;\,\mathbf{C}=0.5772\,$}
    &+&+ \\ \hline
\{3\}&$(\ln r)^n\,;\:${\scriptsize $n\geq 2$}&{\small$(\ln Q^{-1}-
      \mathbf{C})^n + \Delta_n$} &+&+ \\ \hline
\{4\}& $\frac{r^2}{r^2+a^2}$& $e^{-aQ}\:$;\,\,{\scriptsize $a\geq 0$}
    &+&+ \\ \hline
\{5\}&~$ e^{-ar}$; {\scriptsize $a\geq 0$}&$\frac{2}{\pi}\,
    \arctan\left(  \frac{Q}{a}\right)$ &--&-- \\ \hline
\end{tabular} \end{center} \smallskip

 In two right columns we mark the correspondence of the function $f(r)\,$
at $r\to\infty\,$ to the Tauberian condition (S), as well as the fulfillment
of condition (QMC) only in the IR region in the form (\ref{qm-corr}). Note,
that symbols  "+" and "-" turn out to be completely correlated. \par

 As it follows from the Table, in accordance with eq.(\ref{tauber}), the class
of admissible functions is rather narrow. For instance, it does not contain
trigonometric functions and exponentials $\sim e^{ar}\,.$ Only the first--line
expression $F_{\{1\}}\,$ corresponds (up to a factor!) to the (QMC) rule. The
second line's one $\,F_{\{2\}}=\ln(Q^{-1})-\mathbf{C}\,,$ (with $\mathbf{C}\,$
being the Euler constant) satisfying the Tauberian condition as $Q\to 0\,,$
generally differs from $f_{\{2\}}(1/Q)\,,$ especially in the region
$Q\simeq 1\,.$ The same is true for the line \{3\}; here the constant
$\Delta_2\,$ had been introduced before in eq. (\ref{lambda-e}). \par
 \begin{figure}[ht]
 \noindent \centerline{\epsfig{file=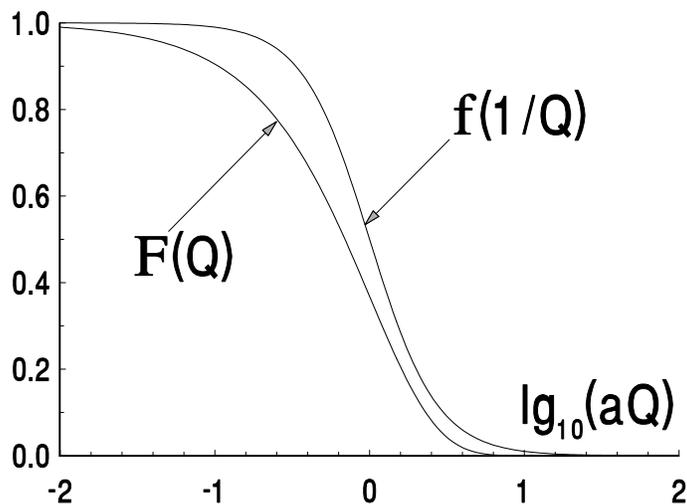, width=90mm}}
 \caption{\footnotesize Behavior of the ``QMC--substituted" function
 $f_{\{4\}}(1/Q)=(1+a^2Q^2)^{-1}\,$ and of its sine-Fourier transform
 $F_{\{4\}}(Q)=\exp(-aQ)\,$.} \end{figure}

The next line \{4\} provides us with an example that satisfies the Tauberian
condition (S), but severely violates the (QMC) rule for the whole function
$\,F_{\{4\}}.$ It is instructive to compare the behavior of the Fourier image
$\,F_{\{4\}}(Q)=\exp(-aQ)\,$ with the ``QMC--substituted" initial function
$f_{\{4\}}(1/Q)=(1+a^2Q^2)^{-1}\,.$ As it can be seen in Figure 1, the
relative error being at $aQ\geq 0.2\,$ of an order of 15\%, quickly increases
and reaches the level of 30\% at $aQ\simeq 1\,.$ \par
 Only last example $f_{\{5\}}\,$ does not satisfy the Tauberian condition,
while the corresponding limiting values $f_{\{5\}}(\infty)=F_{\{5\}}(0)=0\,$
coincide.\medskip

 To give some example more close to the realistic QCD case, consider the
sine-Fourier transformation for the class of functions $f(Q)\,$ that
satisfy the K\"allen--Lehmann spectral representation
\beglab{sin-f}
f(q)\to F(r)=\frac{2}{\pi}\,\int^\infty_0\frac{d\,Q}{Q}\,\sin(Qr)\,
\int^\infty_0\frac{d \sigma\, \rho(\sigma)}{\sigma+Q^2}\,\eeq    
with the density $\rho(\sigma)\,.$ Changing the order of integration
and performing the integration over $Q\,$ with the help of line {\{4\}}
from the Table we arrive at
\beglab{sf-spec}
F(r)=\int^\infty_0\frac{ \rho(\sigma)\, d\sigma}{\sigma}
\left(1-e^{-r\sqrt{\sigma}}\right)\,.\eeq                       

  In the case of positive spectral density, this expression defines a
monotonically rising function of $r\,$ with possible singularity as
$r\to\infty\,.$ \par \smallskip

  As  an explicit example, we consider the analyticized invariant QCD
coupling $\albar(Q^2)\,,$ eq.(\ref{alphan}) with the spectral density
$\rho_{\rm APT}(\sigma)\sim [(\ln\sigma)^2+\pi^2]^{-1}\,$ taken from the
one--loop APT\cite{prl97}. For $a_{\rm APT}=\beta_0\alpha_{\rm APT}$ this
yields{\small
 \beglab{apt}
a_{\rm APT}(r)=\int^\infty_0\frac{d\sigma(1-e^{-r\sqrt{\sigma}})}{\sigma[(\ln\sigma)^2+\pi^2]}=1-I(r)\,;\quad I(r)=\int^\infty_0
\frac{e^{-r\sqrt{\sigma}}\,d\sigma}{\sigma[(\ln\sigma)^2+\pi^2]}\,.\eeq
}
 The r.h.s. integral $I(r)$ resembles the Ramanujan ones\cite{htf}
$$
R(r)\equiv\int^\infty_0\frac{e^{-rx}\,d x}
{x[(\ln x)^2+\pi^2]}=2\int^\infty_0\frac{e^{-r\sqrt{\sigma}}\,
d\sigma}{\sigma[(\ln\sigma)^2+4\pi^2]}= \nu(r)-e^{r}\,$$
expressible in terms of the special transcendental function $\nu(x)\,.$
 We shall use this proximity for analysis of the $I(r)\,$ behavior as
$r\to 0\,.$ The function $\nu\,$ asymptotics is well known \cite{htf}.
One has
\beglab{xx0}
R(r)\to-1+\frac{1}{\ln r^2}+O(\frac{1}{\ln^2 r^2})
\quad\mbox{as} \quad r \to 0\,. \eeq
 The difference
$$
\Delta(r)=I(r)-\frac{R(r)}{2}=3\pi^2\int^\infty_0\frac{d\sigma}
{\sigma}\,\frac{e^{-r\sqrt{\sigma}}\,}{[(\ln\sigma)^2+\pi^2]
[(\ln\sigma)^2+4\pi^2]}$$
being positive, vanishes $\Delta(\infty)=0\,$ at infinity.
At the same time, $\Delta(0)=1/4\,.$ Hence
\beglab{ss}
a_{\rm APT}(r)\to\frac{1}{\ln r^2}\ \quad\mbox{as}\quad r \to 0
\quad\mbox{and}\quad a_{\rm APT}(\infty)=1\,. \eeq                    

  It is instructive to compare $\alpha_{\rm APT}\,$ with the monotonous
initial function
$$
\bar{a}_{an}(Q^2)=\beta_0\albar(Q^2)=\int_0^{\infty}\frac{\rho_{\rm
APT} (w^2)dw^2}{w^2+Q^2}=\frac{1}{\ln(Q^2)}+\frac{1}{1-Q^2}\,,$$
\beglab{z}
a_{an}(0)=1\,; \quad a_{an}(Q^2) \to \frac{1}{\ln Q^2}\,\quad
\mbox{as} \quad Q^2\to\infty\eeq                                 
in the coordinates $r=1/Q\,.$
\begin{figure}[h]
\noindent \centerline{\epsfig{file=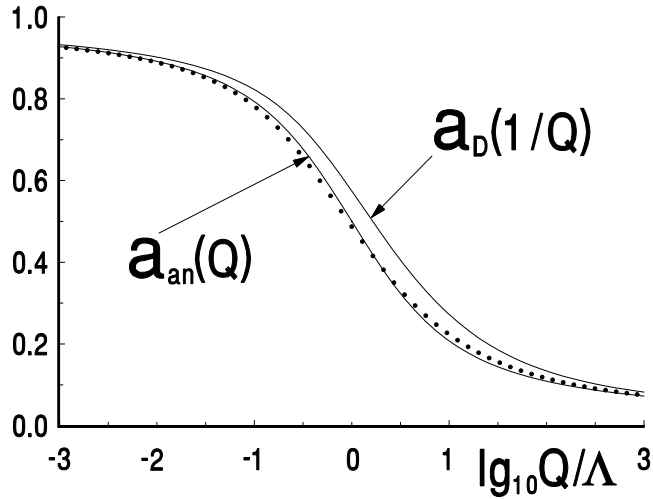, width=85mm}}
\caption{\footnotesize Behavior of the APT--analyticized $a_{an}(Q)\,$ and of
its Fourier transform $a_{\rm D}(1/Q)\,$ with the argument modified by the
quantum-mechanical rule" (QMC) and of its Fourier image $a_{\rm D}(c_F/Q)\,$
(dotted line) with the scale modified by (17).} \end{figure}

Such a comparison (see Fig. 2) reveals a puzzling similarity of both the
functions $a_{\rm D}(1/Q)\,$ and $a_{an}(Q^2)\,,$ that could be essentially
improved by changing $a_{\rm D}(1/Q)\,$ for $a_{\rm D}(c_F/Q)\,.$ In
particular, the relative error in the region $4\lesssim Q/\Lambda\lesssim
70\,$
is reduced from 30\% to 10\%. This produces an impression that one has a
strong argument for supporting the use of the correspondence rules (QMC) and
(\ref{qscale}) for the qualitative estimate of Fourier image of functions
with an ``adequate" --- to eq.(\ref{tauber}) --- asymptotic behavior.
 Unhappily, this impression is erroneous. For example, the analysis of the
family of functions
$$
f(q^2)=\frac{1}{\ln^2(q^2) + b^2}\,;\quad q^2=Q^2/\Lambda^2\, $$
with adequate asymptotics reveals\footnote{Private communication by Dr. A.V.
Nesterenko.} that similarity like in Figure 2 represents a ``puzzling
exclusion" of usual case, which, typically, is far even from very approximate
similarity like the one in Figure 1.\par

Hence, the (QMC) rule, generally, does not provide us with some reasonable
means for qualitative estimation of the Fourier image behavior in the region
far from the asymptotic one even for functions satisfying the Tauberian
condition. \par

\subsection{The ``Schr\"odinger functional QCD coupling" \label{ss3.3}}
  To analyze by nonperturbative means the infrared behavior of QCD, the
``ALPHA collaboration" uses the functional integral approach (both in the
quenched QCD version and with two massless flavors). It works with the {\it
Schr\"odinger functional} (SF) defined in the Euclidean space--time manifold
in a specific way: all three space dimensions are subject to periodic boundary
conditions, while the ``time" one is singled out --- the gauge field values
on the ``upper" and ``bottom" lids differ by a phase factor with the parameter
$\eta\,.$ Then, the renormalized coupling \alphasf is defined via the
derivative $\Gamma'=\partial\Gamma/\partial\eta\,$ of the effective action
$\Gamma=\alpha^{-1}\Gamma_0+\Gamma_1+\as\Gamma_2+\dots$ as (cf. eq.(8.3) in
ref.\cite{lu99}) a function in {\it the distance representation}
$\alphasf(L)=\Gamma'_0/\Gamma'\,$ with $L\,$ being the spatial size of the
above--mentioned manifold.\par

 Quite fresh results reveal the steep rise of the SF $\alphasf(L)\,$ coupling
with $L\,$ in the region $\albarsf\simeq 1\,.$ Here, the analytic fit
\cite{alf01b} to the numerically calculated behavior of \albarsf has an
exponential form
\beglab{sf-exp}
\albarsf(L)\simeq e^{mL}\,\eeq                                         
with $\,m\simeq 2.3/L_{\rm max}\,$ and $L_{\rm max}\,$ (a reference
distance in the region of sufficiently weak coupling) indirectly defined
via the condition $\albarsf(L_{\rm max}) =0.275\,.$ For a discussion of
the momentum--transfer QCD behavior, in the papers of ALPHA collaboration
the ``quantum--mechanical correspondence" rule is used in the form
$$
  L\to 1/\mu\,\eqno{(QMC`)} $$
with $\mu\,$ practically treated as a reference momentum transfer value.
This way of transition from the distance to the momentum transfer picture is
equivalent to the one discussed above in Section 3.1. It works quite well in
the UV region as far as $\,\sim \ln^{-1}Q^2\,$ behavior is compatible with
the Tauberian theorem. Nevertheless, in some precise consideration, like in
numerical relation between various scales\cite{alf95}\footnote{This important
paper was brought to our attention by Dr. U.Wolff. His friendly assistance is
gratefully acknowledged.}, one should take into account the modification of
scale according to eq.(\ref{qscale}).\par
At the same time, the steep rising function of the type
(\ref{sf-exp}) does not fit to the (S) condition of this theorem. Due to this
the IR (i.e., as $\,Q^2\to 0\,$) behavior needs a more elaborate analysis.

\subsection{On the representation of the ALPHA results in the IR
region \label{ss3.4}}
  Generally, there are two different standpoints for discussing this
transition for numerical simulation results obtained on a
finite lattice of size $L\,.$ \par
  The first, simple-minded one treats the obtained numerical results and
their analytic description (\ref{sf-exp}) as an approximation to some limit
with the continuous function $\,\albarsf(L)\,$ that is subject to an integral
Fourier transformation.  On the other hand, as far as a lattice boundary
condition is a periodical
one, it is possible to use a Fourier series instead of the Fourier integral.\par
In what follows, we shall try to discuss the possible Fourier image
of $\,\albarsf(L)\,$ having in mind both the possibilities.

 To start with the integral Fourier transformation, we represent the
ALPHA  ``distance running coupling" \albarsf by the sum of two terms:
$$
\albarsf(L)= \alpha_{\rm PT}(L)+\alpha_{\rm AL}(L)\,$$
with $\alpha_{\rm PT}\,,$ a perturbation contribution, and
$\alpha_{\rm AL}\,,$ an essentially nonperturbative part. \par

  As far as results of {\it all} nonperturbative calculations --- by both the
numerical lattice simulation ones and solving the Schwinger--Dyson equations
(SDE) --- reveal no traces of unphysical singularities, we, on the one hand,
change the first perturbative term $\alpha_{\rm PT}\,$
for some regular expression $\alpha_{\rm PTF}\,$ (like the ``freezed"
one (\ref{apt}) in the Analytic Perturbation Theory or some other
smooth ones like those emerging from the ``effective massive glueball
model" \cite{sim} or from the SDE solving\cite{as01}). \par
 On the other hand, we approximate the second, essentially nonperturbative,
term by expression of type
 \beglab{alp2appr}
\alpha_{\rm AL}(L)\simeq \alpha_k(L)=C_k\,\frac{\pi}{2}\,
\left(\frac{L}{L_m}\right)^k\,e^{m(L-L_m)}\,.\eeq
It is close to (\ref{sf-exp}) and admits further explicit integration.
Here, $k\geq 1\,$ is small integer, and the $\,m L_m\simeq 2.0\,$ and
$C_k\,$ paremeter values follow from results of numerical lattice
simulations. For example, $C_1\simeq 0.013\,.$ \par
That is
\beglab{alp-appr}
\alpha_{\rm SF}(L)=\alpha_{\rm PTF}(L)+\alpha_k(L)\,. \eeq      

 For the Fourier image of regularized $\alpha_{\rm PT-R}\,$ we assume,
qualitatively, the momentum transfer behavior related by the (QMC) rule
\beglab{al-pt}
\albar_{\rm PT-R}(Q)\equiv \mathbb{F}_{sin}[\alpha_{\rm PTF}](Q)=
\alpha_{\rm PT-R}(1/Q)\,\eeq                                            
with a finite IR limit like in (\ref{z})
\beglab{ptf-ir}
\albar_{\rm PT-R}(0)=C\,;\quad 0< C < \infty\,\eeq
which is supported by the second illustration of Section \ref{ss3.2}
see Figure 2.

To perform transformation (\ref{f-sin}) for $\alpha_k(L)\,,$ one needs
some regularization. To this goal, we shall insert the cutoff
factor\footnote{Strictly speaking, this exponential regularization
corresponds to the {\it generalized Fourier transformation} --- see,
\eg Section 1.3. in ref.\cite{titch}.}
$$
 exp\{(m+\mu)\theta(L-\xi L_m)(\xi L_m-L)\}\,;\quad
\xi=2.5\div3\,,\quad \mu\geq 0\,$$
into the integrand of the r.h.s in eq. (\ref{f-sin}). Here, parameters
$\xi\,$ and $\mu\,,$ regulate ``the place of switching-on" and the intensity
of the cutoff.
\par Then, the ``ALPHA coupling function" in the momentum picture can be
represented as a sum of three terms
\beglab{al-mom}
\albar_{\rm SF}(Q)=\albar_{\rm PTF}(Q)+\albar_{\rm k,reg}(Q)
+\albar_{\rm k,sing}(Q;\mu)\,\eeq                                       
with the first one being defined by (\ref{al-pt}) and two others can be
obtained from expressions (\ref{alp2appr}) taken at $k=1\,$:{\small
\beglab{alQAL}
\albar_{\rm 1,reg}(Q)=\frac{C_1}{L_m}\int_{L_m}^{\xi L_m}dL\sin(QL)
e^{mL}=\frac{C_1}{L_m}\left[f_1(Q,\xi L_m)-f_1(Q,L_m)\right]\,;\eeq     
\beglab{alQsing}
\albar_{\rm 1,sing}(Q;\mu)= e^{m\xi L_m}\frac{C_1}{L_m}\int^{\infty}_{\xi
L_m}dL\sin(QL)e^{\mu(\xi L_m-L)}=\frac{C_1}{L_m}\varphi_1(Q,\xi L_m;\mu)\,\eeq
by appropriate differentiation with respect to $m\,.$ }
 Here,{\small
$$f_1(Q,L)=e^{mL}\,\frac{m\sin(QL)-Q\cos(QL)}{m^2+Q^2}\,, $$
$$\varphi_1(Q,L;\mu)=e^{m L}\,\frac{Q\cos(QL)+
\mu\sin(QL)}{\mu^2+Q^2}
\,.$$}
 In the case of the periodical function with a period related to
$\,\xi L_m\,$ the transformation results in the Fourier series rather than
the Fourier integral and the third term in (\ref{al-mom}) is absent.\par

 The most important qualitative feature of the functions $f_k(Q,L)\,$ and
 $\varphi_k(Q,L; \mu)\,$ is their IR behavior. All the functions generally
 tend to zero linearly with $Q\to 0\,.$ The only exception is the case of
 $\mu=0\,$ when one has a power singularity $\varphi_k(Q\to 0,L;0)\to
 1/Q\,.$\par

 By combining this result with (\ref{ptf-ir}) we conclude that $\albarsf(Q)\,$
in the IR region can have a finite limit or the first order pole (the latter
--- only for the case of integral Fourier transformation), contrary to the
exponential growth $\sim\exp{(m/Q)}\,$ that could be anticipated from some
results of the ALPHA collaboration --- we mean, \eg  transition from
Fir.4 to Fig.3 in Ref.\cite{alf01b}.

On a more general ground, one can argue that the exponential growth of the
QCD coupling with $L$ is the utmost steep possible one. In particular, for
the $\ln\alpha(L)\sim L^\nu\,,\quad\nu>1\,$ regime there is no known
mathematical means for defining a Fourier transformation. Vice versa, for
the $\sim\exp\{m/Q\}\,$ IR asymptotic behavior it is impossible to
construct any Fourier transformation and ``return" to the $L$--picture.

\section{Discussion \label{s4}}
 Our first observation concerns the issue of unphysical singularities in
the QCD effective coupling and related QFT quantities like the propagator
amplitudes. These singularities are inconsistent with integral transformations.
In particular, the well-known unphysical singularity in the region
$\,Q\sim\Lambda\simeq 300 - 400\,\,\GeV\,$ prevents the common QCD perturbative
effective coupling \asQ from a straightforward performing of the integration
procedure necessary for transition to the distance representation by the
Fourier transformation. This gives us an additional theoretical {\it argument
against of existence of unphysical singularities in the physically reasonable
theories} like QED and QCD. Remind here, that {\it all\/} the QCD
nonperturbative calculations --- by both the numerical lattice simulation and
the SDE solving --- reveal no traces of unphysical singularities.\par\smallskip

 The second result based on the analysis of the Fourier transformation
deals with the quantum--mechanical correspondence rule
$$
r\to 1/Q\,\eqno{(QMC)}$$
relating the asymptotic behavior of a function $f(r)\,$ as $r\to\infty\,$
and of its Fourier transform $F(Q)\,$ as $Q\to 0\,.$ It has been confirmed
that this rule, being a reasonable guide for some class of asymptotics (the
power and logarithmic type), has its rigid limits of applicability. \par

  First, even for the function with admissible asymptotic behavior, in the
region not very close to the singularity the (QMC) rule yields only a
qualitative correspondence, as it follows from our Figure 1. \par
  Second, it is not valid at all for wide class of asymptotic behaviors
violating the so--called Tauberian conditions\cite{vdz86}, like the
exponential ones.\par
 In particular, the exponentially rising long-range behavior of QCD
coupling in the distance representation
$$
\alpha_{\rm SF}(L) \sim e^{mL}\,$$
observed by ALPHA collaboration on the basis of the lattice simulation of
the Schr\"odinger functional, according to our analysis can correspond in the
momentum (transfer) picture to the \par\bigskip

 a) finite  \ or \hspace{30mm}  b) ``slightly singular"  $\sim 1/Q\,$
 \hspace{30mm} (ALPHA--IR)\par\bigskip
 \noindent IR asymptotics.\par

 This means that these long--distance results, being properly translated to
 the IR region, qualitatively, will not be so far from the results of other
groups\footnote{See, \eg refs.\cite{as01,skw01,bouc01} and short discussion
of their difference in \cite{dv02}.} that perform lattice simulation
calculations (partially supported by solution of appropriate truncated
Schwinger--Dyson equations) for the functional integral defined in the
momentum representation.\par

 From the physical point of view, in our opinion, there are at least
two issues that should be mentioned in connection with the QCD infrared
asymptotic behavior.\smallskip

 First, we have to remember that the region of $Q\lesssim 500\,\MeV\,,$
physically, corresponds to distances $r\gtrsim 10^{-13} cm\,,$ that is to
the hadronic scales. Here, all the QCD notions, like gluonic and quark
propagators, seem to be meaningless.
Even in a more moderate region $0.5\,\MeV\lesssim Q\lesssim 3\,\GeV\,$
of strong QCD interaction there arises a question of physical meaning of
nonperturbative QCD functions, including the effective coupling one. \par

  The mentioned above results for the effective QCD coupling
obtained\cite{as01,skw01,bouc01} by a numerical lattice simulation of the
path QCD integral in the momentum representation are formulated by the
``common QFT language" using the vertices with fixed dynamics, like, \eg
in \cite{skw01,sk02} for the QCD model with two massive quarks. There, an
invariant coupling $\,\gbar(Q^2)\,$ is defined on the basis of the
gluon--quark vertex $\Gamma_{q-gl}(0;Q^2,Q^2)\,$ in the particular MOM
scheme with gluon momentum equal to zero. The invariant coupling function
thus defined suffers from the usual drawback of MOM schemes --- the gauge
dependence. Nevertheless, it can be, in principle, considered as
directly corresponding to some definite physical situation (being
incorporated into a series for some observable with gauge--dependent
coefficients). \par \smallskip

   Second, it seems to be reasonable to relate the IR behavior of the
QCD functions with the confinement phenomenon. Here, it is possible to
appeal to the so--called Kugo--Ojima condition\cite{KuOj79} that, physically,
corresponds to the absence of ``open colour" in the asymptotic states.
In the QCD language, this yields the vanishing of the gluon and quarks
fields renormalization constants, that, in turn, is equivalent to the
zero IR limit of corresponding QCD propagators. Such a behavior has
been observed in the most of the ``lattice--simulation QCD
papers"\cite{as01,skw01,bouc01}. Quite recently it has been supported
by Orsay group\cite{bouc02} on the basis of an instanton liquid model
and, in a sense, by analysis\cite{brod02} of the $\tau\,$ decay data.\par
  In this context, the (ALPHA--IR,a) IR behavior seems to be a quite
reasonable possibility to correlate all above-mentioned lattice simulation
results with the each others and with the physics of confinement.\bigskip

{\bf\large Acknowledgements} \medskip

The author is indebted to I.Ya.\,Aref'eva, Yu.N.\, Drozzhinov, S.V.\,
Mikhailov, M.\,M\"uller--Preussker, O.\,P\`ene, A.A.\,Slavnov, V.S. \,
Vladimirov and B.I.\,Zavialov for valuable advice as well as to A.V.
Nesterenko for stimulating discussion and help in numerical calculations.
 This research has been partially supported by grants of the Russian
 Foundation for Basic Research (RFBR projects Nos 02-01-00601 and
 00-15-96691) and by CERN--INTAS grant No 99-0377.

\addcontentsline{toc}{section}{~~~References} %

\end{document}